\documentclass[fleqn,usenatbib]{mnras}

\usepackage{newtxtext,newtxmath}
\usepackage[T1]{fontenc}

\DeclareRobustCommand{\VAN}[3]{#2}
\let\VANthebibliography\thebibliography
\def\thebibliography{\DeclareRobustCommand{\VAN}[3]{##3}\VANthebibliography}

\usepackage{graphicx}

\title[Deep Learning nearby peculiar velocities]{Deep Learning nearby galaxy peculiar velocities}
\author[Quigley, Hori and Croft]{
Kevin M. Quigley,$^{1,2}$
Samuel Hori$^{1,2}$
and Rupert A.C. Croft$^{1,2}$
\\
$^{1}$McWilliams Center for Cosmology, Department of Physics, Carnegie Mellon University, Pittsburgh, PA 15213, USA
\\
$^{2}$NSF AI Planning Institute for Physics of the Future, Carnegie Mellon University, Pittsburgh, PA 15213, USA
\\
}

\date{Accepted XXX. Received YYY; in original form ZZZ}
\pubyear{2021}

\begin{document}
\label{firstpage}
\pagerange{\pageref{firstpage}--\pageref{lastpage}}
\maketitle

\begin{abstract}
We explore how information in images of nearby galaxies can be used to estimate their distance.
We train a convolutional Neural Network (NN)
to do this, using galaxy images from the Illustris simulation.
 We show that if
the NN is trained on data with random
errors added to the true distance (representing
training using spectroscopic redshift instead of actual distance), then
the NN can predict distances in a test dataset with greater accuracy than
it was given in the training set. This is not unusual, as often NNs are
trained on data with added noise, in order to increase robustness. In this
case, however, it offers a route to estimating peculiar velocities of nearby
galaxies. Given a galaxy with a known spectroscopic redshift one can use the
NN-predicted distance
to make an estimate of the peculiar velocity. Trying this
using relatively low resolution (1.4 arcsec per pixel) simulated galaxy images
we find fractional RMS distance errors of $7.7\%$ for galaxies at a mean distance of 75 Mpc from the observer, leading to RMS peculiar velocity errors of $440 $km/s.
In a companion paper
we apply the technique to 145,115 nearby galaxies from the NASA Sloan Atlas.

\end{abstract}

\begin{keywords}
methods: data analysis -- techniques: image processing -- galaxies: distances and redshifts -- 
\end{keywords}

\section{Introduction}






Measurements of galaxy distances can be made  with many different techniques,
with the first being Hubble's identification of cepheids as standard candles
\cite{hubble29}. Some reviews of distance determination in cosmology are: \cite{rowan85,freedman10,sha21}. While
some techniques combine spectroscopy and photometry (such as the
Tully-Fisher relation (\citealt{tully77}) to generate rungs on the cosmic
distance ladder, other techniques focus on the information in galaxy
images. An example of the latter is the Surface Brightness Fluctuation
technique (\citealt{tonry88}), which makes use of the
granularity caused by individual bright stars. These fluctuations are
smaller in galaxies at greater distances when more stars are averaged over per
resolution element. Beyond these small scale effects, there is other
information in galaxy images, related to their complex morphologies. For
example the thickness of spiral arms, or the gradients of surface brightness
profiles could yield clues to galaxy sizes in physical units and then enable
their use as a standard ruler. In the present paper we explore the use of
artificial Neural Networks (NN, see e.g. \cite{goodfellow16})
to extract this type of
information, using simulated
data to train and test how galaxy images can be used to estimate galaxy
distances. We will restrict ourselves to nearby galaxies (within $\sim 100$
Mpc), as in this case the relative distance errors will be small enough that
the addition of spectroscopic redshifts will allow estimates of galaxy
peculiar velocities.

When in multiple wavelength
bands, galaxy images and their photometry can also be
used to determine photometric redshifts. Photometric redshifts have
long been established as a valuable counterpart to spectroscopically
determined redshifts (\citealt{baum62},
\citealt{puschell82}), particularly for dim or distant galaxies for
 In addition, the
ever-increasing volume of catalogued galaxies, particularly in
upcoming surveys (such as the Rubin Telescope's Legacy Survey,
\citealt{ive2019}), makes universal spectroscopic analysis
impractical (\citealt{dahlen13}).  While in the past limitations in
the accuracy of photometric redshifts have made the method
insufficient for many scientific uses (\citealt{hogg98}), advances in
techniques such as broadband photometry (\citealt{connolly95}) and
Spectral Energy Distribution fitting (\citealt{bolzonella00}) have
made photometric redshift measurement useful for objects too
faint for spectroscopic redshift determination
(\citealt{hildebrandt10}).  Even still, given the accuracy of
photometric redshifts required for current analysis, such that in dark
energy (\citealt{albrecht06}) or the mass function of galaxy clusters
(\citealt{peacock06}), there remains room for improvement.

It has been realised by multiple authors (e.g., \citealt{collister04}, 
 \citealt{carliles10}, \citealt{momtaz22}), that Machine Learning (ML,
such as NNs) can be a useful tool for estimation of photometric
redshifts.  Already, NNs have proven valuable in several analysis
problems in astronomy, such as point and extended-source
classification (\citealt{atkinson18}), the detection of high-redshift
QSOs (\citealt{yeche10}), galaxy morphology prediction
(\citealt{dieleman15}), and even the determination of galactic
spectroscopic redshifts (\citealt{stivaktakis20}). NNs have been presented
with galaxy fluxes in different wavelength bands and used as an alternative
to traditional template fitting to estimate photometric redshifts (\citealt{collister04}).
Whole galaxy images at multiple wavelengths have also been passed to ML algorithms,
in order to access more information (\citealt{wilson20}). In these cases  it has been
shown that accuracy can increase with respect to traditional methods
(such as template pre-calculated band magnitudes and colors, see e.g., \citealt{hoyle15}) and more traditional techniques like Spectral Energy Distribution fitting (\citealt{tagliaferri03}), having
been applied to galaxies with redshifts between $z=0$ and $z=2.0$.

In the present work, we apply similar techniques to relatively nearby galaxies, in order to estimate
distances (instead of photometric redshifts), and hence peculiar velocities.
Rather than training our network on already-calculated photometric values such as overall broadband filter fluxes or half-light radii, we instead train the NNs on multi-band images of entire galaxies.
By presenting entire galaxy images to the NNs, rather than just summary statistics which we expect to be relevant for galaxy size and luminosity determination, we hope both to avoid imposing prior convictions about what photometric data should be relevant in analysis, and for the networks to perform more advanced analysis on the images themselves, including:

\begin{itemize}
\item{Calculating galactic luminosities in each band, therefore calculating colour}
\item{Identifying structure within galaxies and perhaps even galaxy type}
\item{Analyzing the effect of surface brightness fluctuations, which have been shown to be an indicator of distance (\citealt{tonry88})} 
\item{Finding relevant photometric features which are not yet recognized to be correlated with galaxy size or luminosity}
\end{itemize}

Our plan for the paper is as follows. In Section \ref{nn} we briefly describe the Machine Learning technique we use in our analyses. In Section \ref{method} we describe the simulated galaxy data, how we pre-treat it and how we train the  NN  and use them to predict
galaxy distances. Our results are presented in Section \ref{results}, including the
predictions for galaxy peculiar velocities. We discuss the results and possible future
work in Section \ref{discussion}. In Appendix \ref{appendix_a} we give some details
of the Machine Learning architectures and parameters and their relative performance.

\section{Neural Networks and their architectures}
\label{nn}
Neural Networks (see the introduction by \citealt{goodfellow16}) have become an indispensable tool for classification and regression problems, provided that large samples of pre-classified training data can be provided.

A neural network consists of layers of "neurons."  Each layer takes a linear input of neurons in previous layers before applying a non-linear transformation.  Parameters of the network are then typically trained to produce accurate results on a test set through the backpropagation algorithm. 
Previous work has shown a variety of neural network architectures to generalize well to problems of interest. 

One architecture frequently used for structure identification in images is the Convolutional Neural Network (CNN, \citealt{lecun98}, \citealt{krizhevsky12}).  CNNs typically take as input collections of multi-channel images, with each channel in each pixel serving as a single input neuron.  The first hidden layer of neurons performs two-dimensional convolutions on each input channel, returning several convolved images.  A second hidden layer may in turn convolve those images again, returning a second intermediate collection of images, and so on.  CNNs explored in this paper consist of between two and five convolutional layers followed by a "fully-connected" output layer, in which each remaining neuron feeds into a single output, trained to represent the network's final prediction for galaxy distance.

The other main architecture explored in this paper is the Residual Neural Network (ResNet, \citealt{he16}), which is a variant of a CNN.  Rather than simply feeding the output of each hidden layer sequentially into the next convolution, the ResNet architecture also directs the outputs of some layers into later nonconsecutive layers.  ResNet architectures have been applied widely in computer vision and generally been shown to outperform ordinary CNNs.  As with the CNNs, the ResNets explored in this paper have a fully-connected output layer, with network optimization to be treated directly as a regression problem. 

\begin{figure}
    \includegraphics[width=1.0\columnwidth,trim=2.0cm 1.5cm 1.5cm 1.5cm,clip]{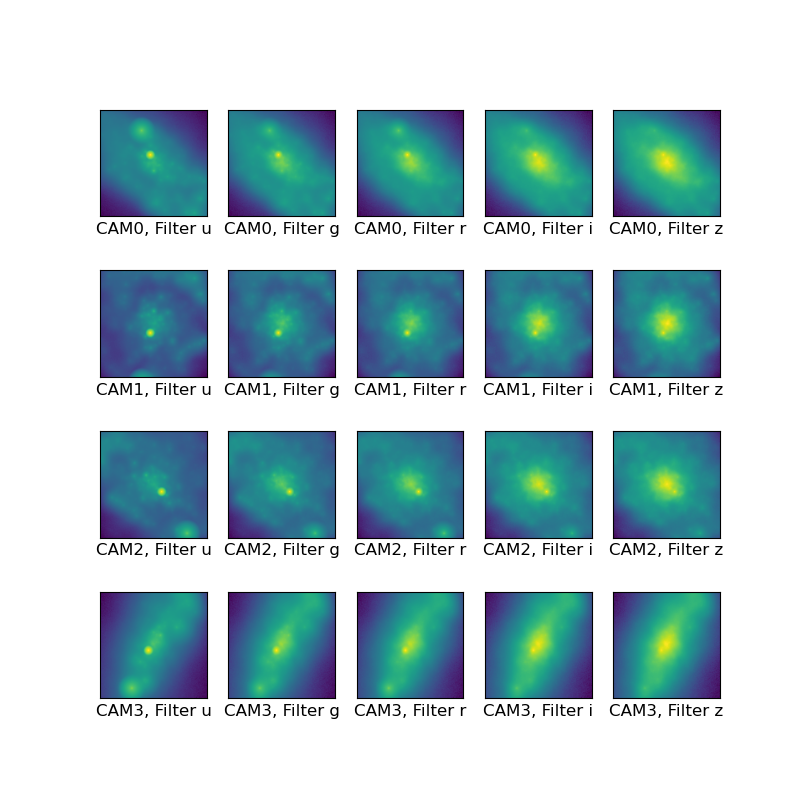}
    \caption{Example raw image data for a single typical subhalo, presented on a logarithmic brightness scale.  Each of the four rows correspond to a different view angle from the Illustris public dataset, and the five columns correspond respectively to the $u,g,r,i$ and $z$ filters of each image.  Since each image was presented to the NNs as a separate input, the above figure illustrates four different elements in our dataset.}
    \label{fig:allcams}
\end{figure}

\begin{figure}
    \includegraphics[width=\columnwidth,trim=2.0cm 0.5cm 1.5cm 1.5cm,clip]{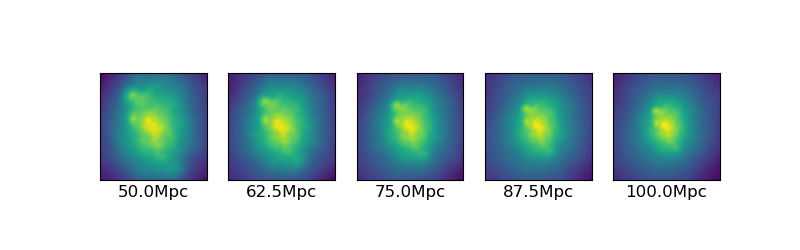}
    \caption{Examples of scaled image data for a single subhalo simulated to be appear at several different distances from the observer, presented on a logarithmic brightness scale.  The above image represents only data from the $z$ filter, but all filters for each image were scaled identically.}
    \label{fig:allsizes}
\end{figure}

\section{Methods}
\label{method}

\subsection{Dataset Retrieval and Preparation}
Training images were retrieved from the public dataset of the Illustris Project\footnote{\tt https://www.illustris-project.org/data/ }, a set of cosmological simulations run to explore galaxy formation and large-scale structure (\citealt{vogelsberger14}).  All data used were images retrieved from the Illustris-1 simulation at redshift $z=0$ (\citealt{nelson15}).  Queries were made for every primary-flagged subhalo for which the Illustris collaboration publicly provided a \textit{.fits} file, limiting the dataset to galaxies with masses above $10^{10} M_\odot.$\footnote{\tt https://www.illustris-project.org/data/docs/specifications/}

Each of the 4,642 galaxies queried had four associated images, taken at the corners of a tetrahedron centered on the galaxy's potential minimum with a uniform distance of 50 Mpc (\citealt{torrey15}).  Each image is  square, consisting of $256\times256$ pixels,
corresponding to a physical scale of $84.4$kpc. The
fiducial distance of 50 Mpc was used to set the angular size of each pixel, $1.36$ arcsec.
Raw data for each image consisted of 36 filters, of which only the SDSS $u, g, r, i, $ and $z$ filters were used in training.  This was done for the sake of imitating true observational data found in the NASA-Sloan Atlas,\footnote{\tt www.nsatlas.org} the target dataset in a future companion paper, Hori et al. {\it in prep.}.  After retrieval, each image was treated as a separate entity, quadrupling the size of the dataset to 18,568 five-filter images.  We give an example of raw image data used in Fig.~\ref{fig:allcams}.

Before being presented to the NNs, all images were altered to simulate a range of galaxy distances.  First  a random 'target' distance was assigned to each image to be simulated, the distributions of which varied across trials (see Section \ref{dataparams}).  Next, image resolutions were scaled so that the angular field of view of the image---and therefore galaxy solid angle---was consistent with the new galaxy distance.  This transformation was accompanied with a bicubic interpolation.  Images were then center-cropped to $128\times 128$ pixels to maintain uniform image resolution and angular field-of-view across the dataset.  Finally, all pixel brightnesses were scaled to preserve absolute galaxy luminosity, using the inverse-square law appropriate to the Euclidean space approximation for nearby objects.  Image filter bands were broad enough that the relatively small Hubble flow redshift changes would only negligibly alter galaxy color (\citealt{snyder15}), so no spectroscopic edits were made to the image data.  Because of this, the NNs worked exclusively with visual information to ascertain galaxy distance, rather than spectral information.  We give an example of scaled image data in Fig.~\ref{fig:allsizes}.

In order to simulate the most common observational situation, where galaxy redshifts would be available to train against rather than distances,
in some dataset preparations, further edits were made to simulate galactic peculiar velocity.  In these trials, the distance associated with each image for the NN to train against was modified without transforming the corresponding image.  As a result, Hubble flow redshift was changed without a corresponding change in spectroscopic information.  The resulting difference between redshift and Hubble redshift is then galactic peculiar velocity.  Like Hubble flow redshift at these distances, peculiar velocity redshift was not judged significant enough to justify color edits to the image data. 

Finally, the dataset was randomly partitioned into three groups:  a training set, a validation set, and a test set.  NN performance on validation sets was checked throughout the testing process to optimize network hyperparameters, but test sets were only presented to the NNs when generating the final performance results for this paper.  65\% of the original dataset was allocated to training, 10\% to validation, and the final 25\% to the test set.

\subsection{Dataset Parameters}
\label{dataparams}
Simulated galaxy distances were randomly selected between $50$Mpc and $100$Mpc, either from a uniform distribution function between these values or else from a normal distribution with mean $75$Mpc and standard deviation $14.5$Mpc.  This limited distance regime was selected out of concern for transforming the Illustris image data with high fidelity.

Typical peculiar velocities were assumed to be normally distributed about zero with a standard deviation of $400$km/s, approximately consistent with
observational estimates (e.g., \citealt{peacock01}).  Using a Hubble Constant of $70$ kms$^{-1}$Mpc$^{-1}$ (\citealt{abbott21}), these peculiar velocities correspond to target distance changes with
a standard deviation of $\frac{400}{70}\approx 6$Mpc.  For this reason, when present, target distance changes were randomly selected from a normal density function with a mean of $0$ Mpc and standard deviation of $6$ Mpc.  Peculiar velocities were selected independently of galaxy distance.

Normalization of training data has been shown to have the potential to increase NN training speeds and generalization capabilities (\citealt{huang20}), so in some trials, normalization methods were applied to the training data.  In datasets where galaxy distances were selected from a uniform distribution, the training values were shifted and scaled to fall between $0$ and $1$.  When a normal distance distribution was used, however, the mean target distance was shifted to $0$, then all target values uniformly scaled to have standard deviation of $1$.

\subsection{Network Preparation}
\label{netprep}

Using the open-source \texttt{pytorch} package for Python\footnote{\tt{https://pytorch.org}}, we developed and trained several NN architectures to take as inputs the edited Illustris images and return the simulated distances associated with them.  Our investigations focused primarily on CNNs with between two to five layers and ResNets of 18 and 50 layers.  Networks used only L2 Loss for training (we refer to the Mean Squared Error as the L2 loss, although strictly it should be the square of the Mean Squared Error):
\begin{equation}
{\rm L2\ loss}=\sqrt{\frac{1}{N}\sum^{N}_{i=1}(D_{\rm true}-D_{\rm predicted})^{2}},
\label{l2loss}
\end{equation}
though in analysis other bulk performance indicators were consulted as well, including L1 Loss (Mean Absolute Error):
\begin{equation}
{\rm L1\ loss}=\frac{1}{N}\sum^{N}_{i=1}|D_{\rm true}-D_{\rm predicted}|,
\label{l1loss}
\end{equation}
 and Average Shift (Mean Error):
 \begin{equation}
{\rm Average\ Shift}=\frac{1}{N}\sum^{N}_{i=1}(D_{\rm true}-D_{\rm predicted}),
\end{equation}

To optimize the training process, a preliminary array of small NNs were trained for a short time.  This was done in order to probe whether a particular combination of hyperparameters would be immediately apparent as more effective.  Hyperparameter values explored were:

\begin{itemize}
\item Kernel size of first-layer convolutions -  3, 5, and 7 pixels
\item Number of first-layer convolutions - 6, 12, 18, and 24
\item Kernel size of second-layer convolutions - 3, 5, and 7 pixels
\item Number of second-layer convolutions - 6, 12, 18, and 24
\item Convolution strides - 1, 2, and 3 pixels
\end{itemize}

Two-layer CNNs with each combination of the above hyperparameters was trained for approximately five epochs, but no difference greater than $2\%$ in L2 validation loss was observed after training.  We made the inference that within the tested range, hyperparameters involving convolution characteristics were largely irrelevant to overall network performance.  In later testing, then, we placed more emphasis on other aspects of training, such as optimization of learning rate and normalization of target data.

\begin{figure}
    \includegraphics[width=\columnwidth]{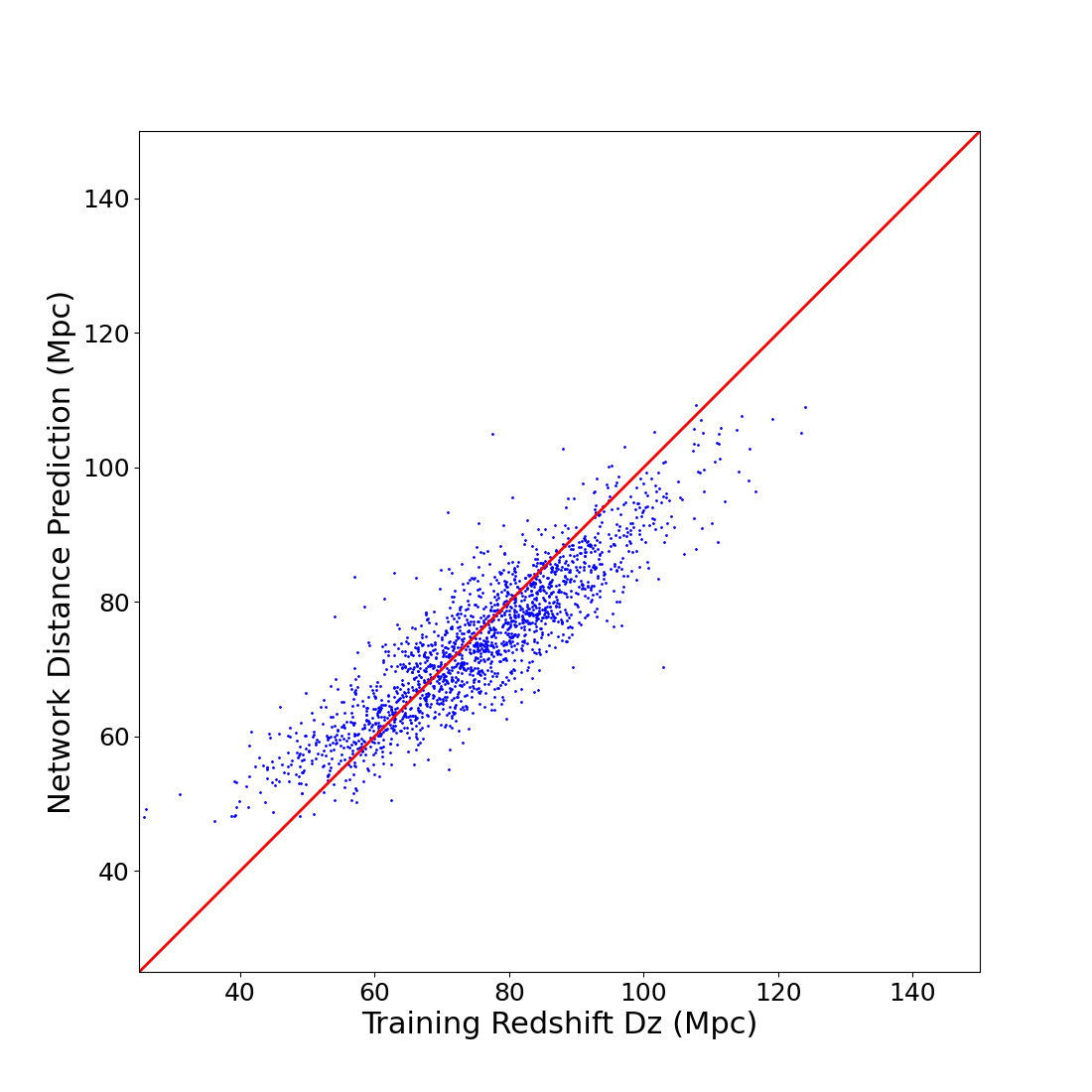}
    \caption{Example of NN outputs after training using galaxy images labelled with distances ($D$, which don't including the effect of peculiar velocities).  Each point represents a single galaxy image from the test set, with simulated distance $D$ on the x-axis and network output on the y-axis.  The red line indicates perfect correspondence.  This data comes from a 3-Layer CNN trained on a normalized dataset with a normal distance distribution.}
    \label{fig:results_noErrs}
\end{figure}

\begin{figure*}
    \includegraphics[width=2.0\columnwidth,trim=3.5cm 1.5cm 3.5cm 3.5cm,clip]{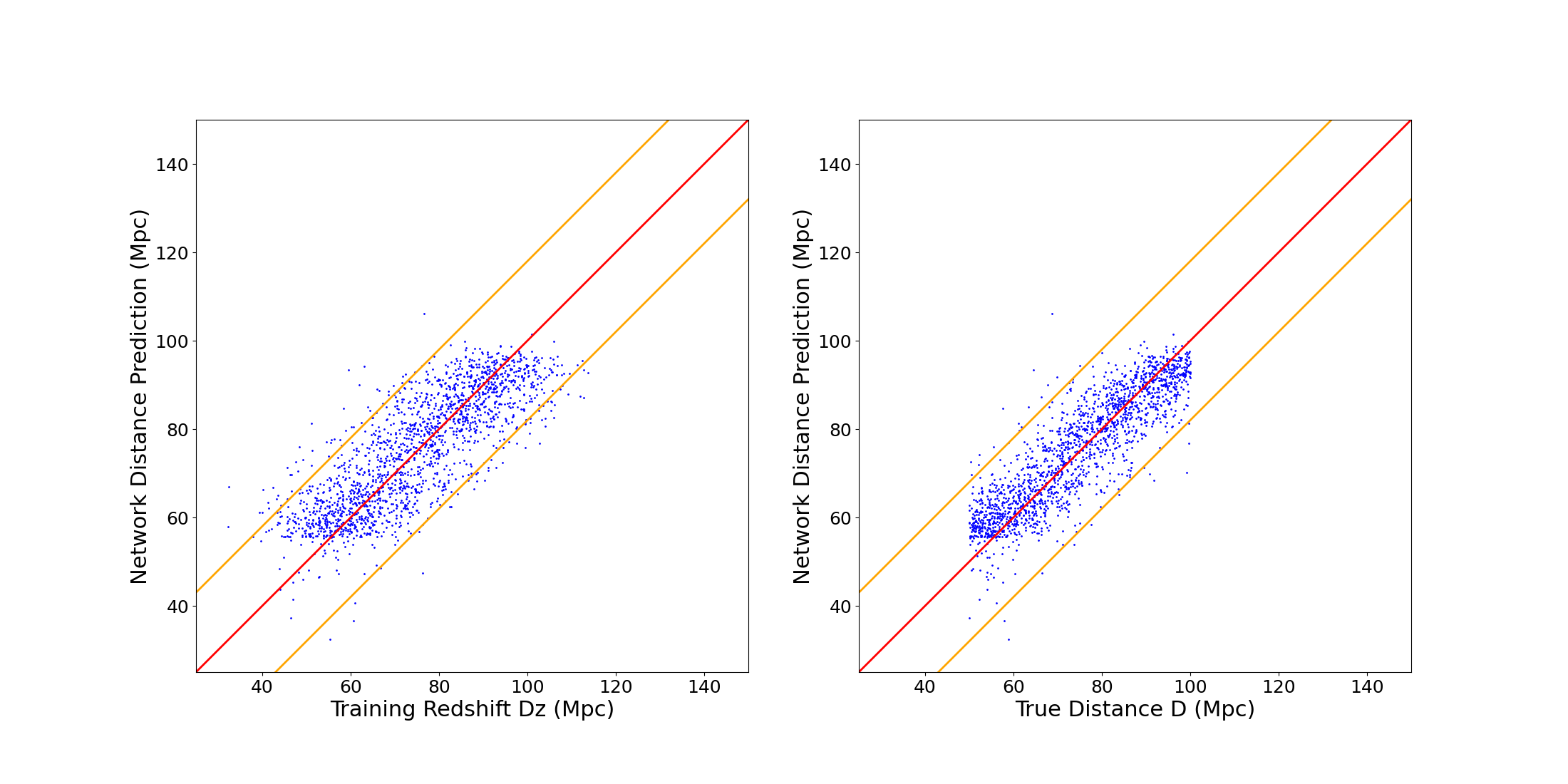}
    \caption{Examples of NN outputs after training using galaxy images labelled with
    "distances" that include the effects of peculiar velocity ($D_z$, defined in Equation \ref{dz}).  Left panel: Each point represents a single image in the test set, with $D_z$
    on the x-axis and network outputs on the y-axis.  Right panel: Each point has true distances $D$ on the x-axis instead.  The red line in each graph indicates a perfect network prediction, and the orange lines (to guide the eye) indicate three standard deviations of the parent distribution of peculiar velocity.  The tighter correlation exhibited in the second graph illustrates that L1 Loss between network output and $D$ is lower than that between network output and $D_z$.  This data comes from a 2-Layer CNN trained on a normalized dataset with a uniform distance distribution.}
    \label{fig:results_Errs}
\end{figure*}

\begin{figure}
    \includegraphics[width=\columnwidth]{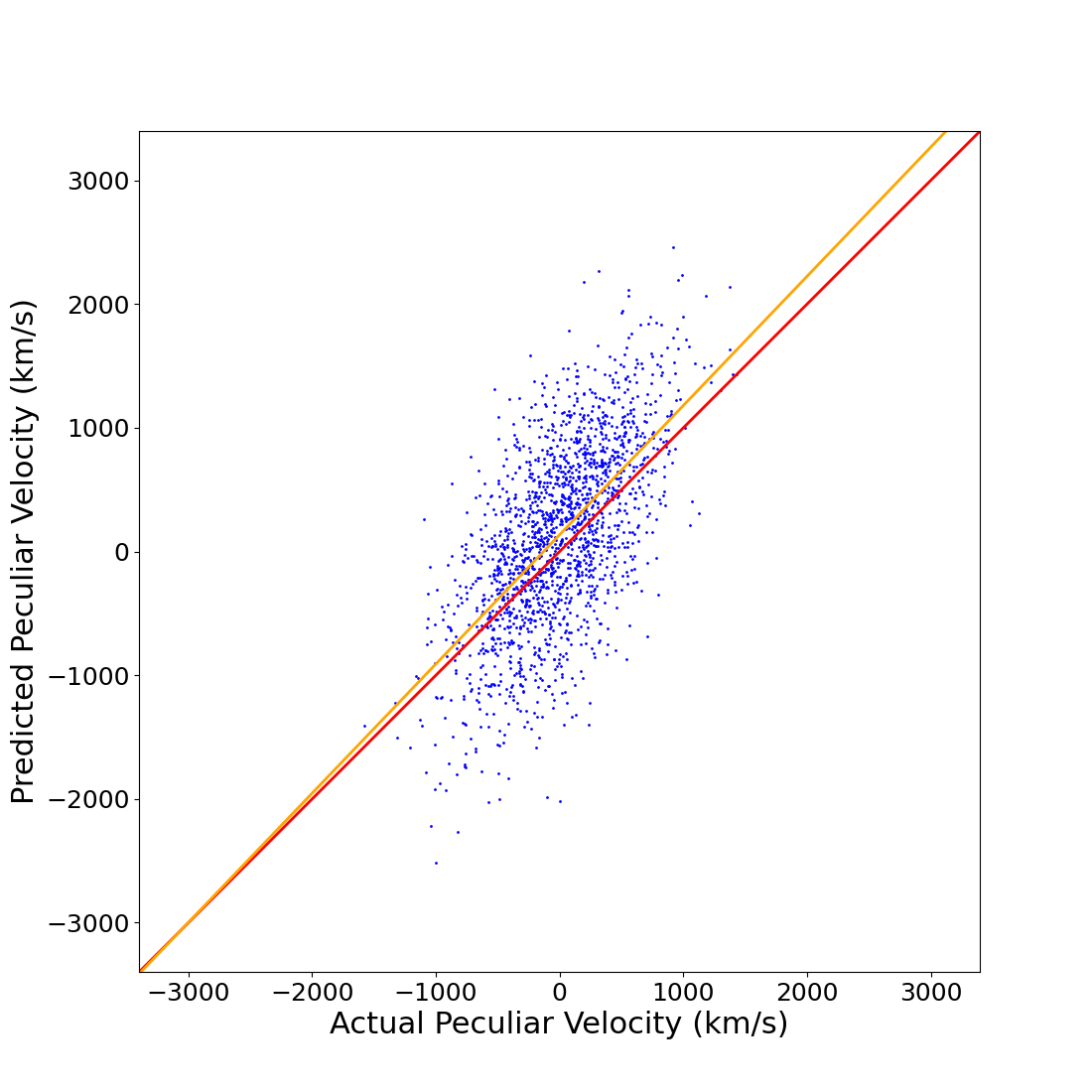}
    \caption{Example of NN output interpreted as a peculiar velocity prediction.  Each point represents a single image in the test set, with the x-coordinate representing its simulated peculiar velocity, and the y-coordinate representing the difference between the neural network's distance prediction and the original data point distance.  The red line indicates perfect correspondence, and the orange line represents the line of best fit.  Apparent distances have been converted to peculiar velocity values using a Hubble constant of $70$kms$^{-1}$Mpc$^{-1}$.  This data is from a 3-Layer CNN trained on a non-normalized dataset with a uniform distribution of galaxy distances, whose line of best fit has a slope of $1.044$ and root-mean-square error of $443.632$ kms$^{-1}$ (see Appendix \ref{appendix_a} for more details of the architectures tested).}
    \label{fig:results_PV}
\end{figure}

\section{Results}
\label{results}


The idea behind our method is that our NNs are trained using distances $D_z$ derived
directly from spectroscopic redshifts, where the conversion assumes Hubble flow only:
\begin{equation}
D_{z}=\frac{cz}{H_{0}}.  
\label{dz}
\end{equation}
Here $cz=v_{p}+H_{0}D$, where $v_{p}$ is the line-of-sight component of the
galaxy peculiar velocity and $D$ is the true distance.
Although $D$ would not
be available in an observational dataset, the NN will nevertheless output an estimate of 
$D$. This is because, according to the cosmological principle, the averaged $D$ and $D_z$ values binned as a function of redshift should be the same. The NN is in essence
 learning the average relationship between galaxy images and their distances. The
 effect of peculiar velocities when the NN is trained using $D_z$ is to act as a
 type of noise. We note that in other contexts, noise is added to NN training data on purpose (see e.g., \citealt{noh17}), to reduce overfitting and to improve robustness.

Although the network has no way of determining directly from an image what fraction of redshift arises from peculiar velocity, we assume that spectroscopic redshifts (and hence $D_z$) will be available (in fact we have assumed that the NN would be trained using these values).  By comparing the NN estimate of the distance to a galaxy $D$ to the observed value of $D_{z}$ , we can obtain a value for peculiar velocity: $v_{p}=H_0 (D_{z}-D)$.  We are in general most interested in the case where the NN was trained using $D_{z}$ but used to predict $D$, allowing us to predict peculiar velocity.  We also, however, compare the NN prediction of $D$ directly to $D_{z}$, as well as first investigating the simplest case, a NN trained using the true distance, $D$.

\subsection{NN Performance trained using $D$}
\label{truedist}

As a first test, we train the NNs on the true simulated galaxy distances $D$, with no introduced peculiar velocity.  When network output is compared directly to this $D$ it gives a measure of how well the networks were able to identify the aspects of galaxy images affecting distances.  In Appendix \ref{appendix_a} we give detailed results and break them down by training dataset and architecture. We show an example of results for such a network trained using true distance $D$ in Fig.~\ref{fig:results_noErrs}.

Over all network trials with peculiar velocities, the minimum error for network outputs compared to true distances $D$ was $4.535$ Mpc L1 Loss (Equation \ref{l1loss}), or $6.04\%$ of the average simulated distance of $75$Mpc.
A large factor in network performance appeared to be whether the associated dataset had been normalized, with networks trained on normalized datasets having an L1 Loss lower than their
non-normalized counterparts by an average of $2.452$Mpc L1 Loss, resulting in an L1 Loss (average absolute error) of $5.051$Mpc, or $6.73\%$ of the average simulated distance.  

Network architecture itself also played a role in accuracy, with CNNs outperforming analogous ResNets by $3.765$Mpc L1 Loss, on average.  Among CNNs, no particular architecture universally performed better than the others, but between the two ResNet architectures, the smaller 18-layer version outperformed the 50-layer model on average by $4.231$Mpc L1 Loss.  
These differences in training quality can largely be attributed to the disparity in training time required for ResNets as compared to CNNs.  While the ResNets did indeed complete their training, the relatively high learning rate may have prevented the networks from adequately approaching a minimum.  Future work could investigate training the ResNets for a much longer time at a lower learning rate, and then it is expected that the performances of the two architectures would be more comparable.

\subsection{NN performance when trained using $D_z$}
\label{performance}

For the other set of tests, we introduce a simulated peculiar velocity to the training data by using $D_z$ (Equation \ref{dz}) in training,  while comparing the NN output to both $D_{z}$ and to the true distance $D$.  The effectiveness of network convergence, therefore, is evaluated by comparing network output to $D_z$, whereas the viability of network output as a predictive tool is evaluated by comparing it to $D$.

Over all network trials with peculiar velocities, the minimum error for network outputs compared to true distances $D$ was $4.901$Mpc L1 Loss, or $6.53\%$ of the average simulated distance of $75$Mpc.  Unsurprisingly, networks trained using the true distance $D$ (Section \ref{truedist}) were more accurate than networks trained using $D_z$.  However, this outperformance was not as pronounced as might be expected, with networks trained using $D$ only outperforming their $D_z$-trained counterparts by an average of $0.164$ Mpc L1 Loss.

Again, normalization appeared to be the largest factor in network performance, with networks trained on normalized datasets having an L1 Loss lower than their non-normalized counterparts by an average of $1.628$Mpc L1 Loss.  Architecture also continued to play a very large role, with ResNets still outperforming CNN counterparts by $6.072$Mpc L1 Loss, on average.  Again, no particular CNN clearly outperformed another, but the 18 layer ResNet outperformed the 50 layer model by $2.675$Mpc L1 Loss.

A crucial finding is that, for all datasets which contained peculiar velocities, the network output more closely modeled the true distance values $D$ than the $D_z$ values that they had been trained against.  L1 Loss against $D$ was on average $1.521$Mpc lower than L1 Loss against $D_z$, with the effect being particularly pronounced in the CNN trials, for which L1 Loss against $D$ was $1.644$Mpc lower than L1 Loss against $D_z$.  
This improvement in performance indicates that rather than simply minimizing the loss functions for the data that had been presented to them, the NNs were instead identifying underlying aspects of galaxy images which indicated true simulated distance, cutting through the randomly introduced noise.  Such behavior gives promise to the use of NNs as a predictive tool for galaxy distance determination, especially in the presence of peculiar velocity redshifts.  We give an example of network results compared to $D_z$ and $D$ respectively in Fig.~\ref{fig:results_Errs}.

\subsection{Network Output as a Peculiar Velocity Prediction}

One of the advantages of the NN technique as tested is that it allows for the determination of peculiar velocities.  After the NN is trained on datasets with redshifts $D_z$, we record the predicted distances output by the NN, $D_p$.  Then, we take the original redshifts $D$ as given values for true galaxy redshift.  By calculating $v_{p, {\rm pred}} = H_0 (D_p - D)$ and using a Hubble Constant value of 70 kms$^{-1}$Mpc$^{-1}$, we  obtain a NN prediction for the peculiar velocity.  By comparing this prediction to the actual simulated peculiar velocity $v_p = H_0 (D_z - D)$, we can evaluate the networks' abilities to discern peculiar velocity.

We find lines of best fit for all graphs of predicted versus simulated peculiar velocity values, using least-squares linear regression.  In all cases, the slope of this line of best fit is found to be very close to one, often much closer than a qualitative inspection might imply (see the example shown in  Figure \ref{fig:results_PV}).  Slopes are always within $5\%$ of one ( the values 
are given for different NN hyperparameters in Table \ref{tab:results}).
The intercepts of the best-fit lines are on the order of $100$km/s, with the majority below $25$km/s.

The minimum root-mean-square error of the NNs peculiar velocity predictions over all trials and datasets is $443.632$ km/s, or just over the standard deviation of the simulated peculiar velocity.  On the whole, neither normalization nor network architecture contribute significantly to the network's ability to predict peculiar velocity.  Instead, the biggest factor determining performance seems to be the distance distribution of galaxies used in training and testing.  Uniformly-distributed training sets exhibit predictions up to $\sim 250$ km/s more accurate than normally-distributed training sets.

\section{Summary and Discussion}
\label{discussion}

\subsection{Summary}
Our results show that neural networks, particularly Convolutional Neural Networks, can be used to obtain estimates of galaxy distance and line of sight peculiar velocity using commonly available data from sky surveys, including spectroscopic redshift and galaxy images in the $u$, $g$, $r$, $i$, and $z$ bands.   We also demonstrate more specific results:

\begin{itemize}
    \item{Given a training set of multiband galaxy images with associated distances, networks were routinely able to predict those distances within $10\%$, with a minimum RMS error of $7.7\%$ or $5.8$ Mpc.} 
    \item{When redshifts (including peculiar velocities) where used to train the NN
   rather than true distances, this did not significantly spoil the networks' ability to predict those distances, increasing the average RMS loss by only $0.164$Mpc, or $2.7\%$.}
    \item{In datasets with simulated peculiar velocity, networks predicted true unaltered distance $D$ more accurately than the altered training distance $D_z$ by an average of $1.628$Mpc L1 Loss, indicating that the network was truly identifying information from the images, rather than memorizing the methods used to create the dataset.}
    \item{When the network distance prediction is used to predict peculiar velocity: $v_{p}=H_{0}(D_p - D_{z})$, there was no observed dependence on $v_p$ error as a function of $v_p$ magnitude, resulting in a line of best fit with a slope very close to one (between $5\%$ and $0.1\%$):  on the average, networks very accurately predicted peculiar velocity.}
    \item{RMS error for network peculiar velocity predictions were as low as $440$km/s, just over the standard deviation of the simulated peculiar velocity.}
\end{itemize}

\subsection{Discussion}

Our tests of whether galaxy image data can be used to infer distance from the observer
and then peculiar velocity are obviously quite crude. We have carried them out as
a proof of concept, but any serious determination of the potential accuracy of
the method must wait for future work, including with actual observed galaxy images. Below we outline some related points.


First, we address the observation that our fractional distance error between network prediction and true galactic distance is not constant.  On the contrary, it is absolute distance error which remains constant, resulting in graphs which are just as tightly clustered at 100Mpc as they are at 50Mpc, as in Fig.~\ref{fig:results_Errs}.  While other analytical methods might expect absolute error on distance predictions to increase as distance magnitude increases, the metric by which we trained our neural networks explains why this is not observed:

Our networks were trained using L2 loss on absolute distance prediction.  That is, the network was trained to minimize the square of the absolute difference between its prediction and the training value.  Therefore, regardless of the actual value of the training distance, the network punished those predictions which were furthest away and rewarded the ones which are closest.  The result is a trained network which has an approximately constant average absolute difference between its prediction and training value, regardless of what that value is.  If the network instead had been trained using fractional error, then we would indeed expect fractional error to be constant and absolute error to vary with distance.

The data from the Illustris simulation, while reproducing many properties of observed galaxies (\citealt{nelson15})
lacks the complexity of true galaxies. Subgrid physics implementations (e.g., \citealt{springel03}) are used to model processes which strongly affect the appearance of galaxy structure, such
as star formation, and supernova winds. The mass and gravitational force resolutions of
the fully cosmological simulations also restrict the details in the images. For example,
the 0.7 Kpc force resolution of Illustris corresponds to a scale of 6 arcsec at 
the 100 Mpc limit of our mock observations. This is a factor of a few larger than the
pixel size of SDSS images (for example in the NASA Sloan Atlas, e.g., \citealt{maller09} which we plan to use in future work), and two orders of magnitude coarser than data from the HST (e.g, \citealt{lee22}). The variety of galaxy substructures that could be seen with much higher resolution is
likely to be greater. Whether this will make the distance measurements more
of=r less accurate is debatable. For example, with a greater variety of structures the training data may be less likely to capture the full range of possible images at each distance
and this would act as a limit on accuracy. On the other hand, other things being equal, more detail and more
information in each image should lead to better distance constraints.
Carrying out resolution tests (for example by smoothing the data and measuring the
accuracy of inferred distances) would be a way to better understand these effects. 

The training in our tests was also carried out on a limited dataset, distributed 
between 50 and 100 Mpc from the observer. The geometry of a true survey would be different,
with the number of objects in the training set increasing with distance. This could
potentially lead to biases, which could be tested with more sophisticated mock surveys.
We have also not included the large-scale structure of the galaxy distribution when carrying out our analysis. In a real survey, galaxies at similar distances are
likely to be clustered along with others in the training set. If there are environmental
effects which cause galaxies in similar environments to have similar internal structure, then this could also cause biases. Environmental effects have been searched for 
in other distance indicators (e.g., \citealt{mocz12,joachimi15}), and more sophisticated analyses should test
and address this possibility here.

The simulated data we have used are images in fixed wavelength bands, but
because of the low redshifts involved ($z < 0.03$) 
we have made the assumption that the rest and observed wavelengths are the same.
If we had modelled the effect of redshifting on the wavelengths, this would have 
lead to slight changes in the observed galaxy images. Previous uses of CNNs to improve
galaxy photometric redshifts using images (e.g., \citealt{collister04}) spanned much larger redshift ranges, and
therefore this effect was important. In our case, as the effect is subdominant to the change in images due to distance, we do not include it here. We note that the peculiar velocity component  as well as the Hubble component of redshift can have observable effects on galaxy selection and galaxy images (\citealt{alam17}).

One of the main features of our method is that the NNs are trained using redshifts rather than distances ($D_{z}$ rather than $D$ in the notation of Section \ref{performance}). In the case of observations, this would be the most practical situation
as accurate redshift independent distance estimates are not available for most galaxies.
For simulated data, however we have also tested training using the actual galaxy distances (Section \ref{truedist}). Comparing to the results of training using redshifts  lets us know how much this limits the accuracy of inferred distances and peculiar velocities. Comparing the $rms$ errors on inferred distances,
we see that training set distances are somewhat useful. In the future we could 
therefore try increasing the accuracy
of the training set using an iterative process. Such a technique would involve a first
round of training using redshifts, after which the the peculiar velocities inferred by the NN could be used together with the redshifts to yield distance estimates. These distance
estimates could be used (after reinitializing the NN weights) for a second round of training, and so on, iterating until convergence. It is not clear whether this would improve the accuracy, but might be worth trying.

Of course using galaxy distance indicators calibrated using redshifts as
we have done is nothing new. For example, the Tully-Fisher (\citealt{tully77})
or D$_{\rm n}-\sigma$ relations (\citealt{djorgovski87,dressler87}) which relate spectroscopic
properties of galaxies to their absolute sizes and luminosities have both been used extensively to make peculiar velocity catalogues (e.g., \citealt{springob14}) and
were discovered using redshifts as a proxy for distance. In the case
of galaxy images one can hope that a wealth of new information,
available from new imaging surveys such as the Rubin Telescope's Legacy Survey (\citealt{ive2019}) or the Roman HLS (\citealt{dore18})
can be exploited by neural networks and could lead to large new catalogues of
peculiar velocities.





\subsection{Data availability}
Processed data and trained neural networks are available on request to the authors. Raw data was made available by the
Illustris collaboration and is available through their website\footnote{\tt https://www.illustris-project.org/data/}.

\subsection*{Acknowledgments}
We thank the Illustris collaboration for making their data publicly available. SH acknowledges funding from a SURG grant provided by Carnegie Mellon University.
RACC is supported by  NASA ATP 80NSSC18K101,  NASA ATP NNX17AK56G, NSF NSF AST-1909193, and the NSF AI Institute: Physics of the Future, NSF PHY- 2020295. 


\bibliographystyle{mnras}
\bibliography{ref}

\bsp

\appendix

\section{Neural network hyperparameter tests}\label{appendix_a}

We present detailed results of our trained neural network architectures, specifying the hyperparameters and training parameters used, L1 and L2 Loss compared against true distance $D$ and trained distance $D_z$ for all architectures and datasets, and the slope and RMS error for the neural network's peculiar velocity predictions for datasets with simulated peculiar velocity.
 
Simulated galaxy distances were selected from one of two distance distributions:  Normal and Uniform.  Normal distance distributions selected galaxy distances from a Gaussian density function with a mean of 75 Mpc and standard devation of 14.5 Mpc, whereas uniform distance distributions used a constant density function with bounds of 50 and 100 Mpc.  Peculiar Velocities, when present, were selected from an independent Gaussian density function with a mean of 0 Mpc and standard deviation of 6 Mpc.

CNNs were trained between 2 and 5 layers for around 20 epochs, with learning rate and hyperparameters provided in Table ~\ref{tab:params}.  ResNet Architectures trained with a learning rate of 0.00001 on normalized datasets for about 40 epochs, largely with default parameters as supplied by \texttt{pytorch}\footnote{\tt{https://pytorch.org/vision/stable/models.html}} with the output layer replaced by a fully connected layer with no activation function and an output size of 1.
Even with a significantly reduced learning rate and large supercomputer emphasis, we were not able to get ResNets to complete their training on non-normalized datasets in a reasonable amount of time.  This is not to say that they did not train, rather that they did not train quickly enough to return acceptable results for this paper.

\begin{table}
	\centering
	\caption{Summary of Convolutional Neural Network hyperparameters.  Hyperparameters aside from Learning Rate were chosen to be as close to default values as possible, but so that convolutions did not reduce images to a single pixel before the final layer.  Initial tests indicated that these hyperparameters would contribute little to final netowrk performance in this range (See end of Sec. \ref{dataparams}).  We aimed to choose the learning rates so that networks neither overtrained (memorized their dataset), nor fell into a false minimum of parameter space, nor took too long to train at all.  With more computing resources, we could have selected a lower learning rate, thereby improving results.}
	\begin{tabular}{lrr|rrr}
		\hline
        Architecture & Learning Rate & Layer & Filters & Stride & Size \\\hline
        Default & - & 1 & 12 & 7 & 3 \\\hline
        2-Layer CNN & 0.0005 & 1 & 12 & 7 & 3 \\
        & & 2 & 12 & 5 & 3\\\hline 
        3-Layer CNN & 0.0005 & 1 & 12 & 7 & 3 \\
        & & 2 & 12 & 5 & 3\\ 
        & & 3 & 12 & 3 & 3\\\hline 
        4-Layer CNN & 0.0005 & 1 & 12 & 7 & 3 \\
        & & 2 & 12 & 5 & 1\\ 
        & & 3 & 12 & 7 & 3\\ 
        & & 4 & 12 & 2 & 2\\\hline 
        5-Layer CNN & 0.0005 & 1 & 12 & 7 & 3 \\
        & & 2 & 12 & 6 & 2\\ 
        & & 3 & 12 & 6 & 2\\ 
        & & 4 & 12 & 4 & 1\\ 
        & & 5 & 12 & 2 & 2\\\hline
	\end{tabular}
	\label{tab:params}
\end{table}

\begin{table*}
	\centering
	\caption{Summary of Neural Network performance across all major architectures and datasets tested.  Dataset details are discussed in Appendix \ref{appendix_a}, and network architecture details are given in Table \ref{tab:params}.  Only networks trained on datasets with simulated peculiar velocity were used to make peculiar velocity predictions.  An example of network predictions for a dataset with no peculiar velocity compared to true distance $D$ is given in Fig. \ref{fig:results_noErrs}, an example of network predictions compared to both training distance and true distance $D_z$ and $D$ for a dataset with peculiar velocity is given in Fig. \ref{fig:results_Errs}, and an example of network predictions for peculiar velocity is given in Fig. \ref{fig:results_Errs}.
}
	\label{tab:results}
	\begin{tabular}{ll|ll|rrrrrrr}
		\hline
		Dataset & Architecture & L1 Loss & L2 Loss & L1 Loss & L2 Loss & Peculiar Velocity & Peculiar Velocity \\
		& & on $D_z$ (Mpc) & on $D_z$ (Mpc) & on $D$ (Mpc) & on $D$ (Mpc) & Slope & RMS Error (km/s) \\\hline
		Normal Distribution & 2-Layer CNN & 7.482 & 9.514 & 7.482 & 9.514 \\
	    No Normalization & 3-Layer CNN & 7.976 & 10.055 & 7.976 & 10.055 \\
	    No Peculiar Velocity & 4-Layer CNN & 9.300 & 12.522 & 9.300 & 12.522 \\
	    & 5-Layer CNN  & 8.847 & 10.760 & 8.847 & 10.760 \\\hline
	    Normal Distribution & 2-Layer CNN & 4.878 & 6.527 & 4.878  & 6.527  \\
	    Normalized & 3-Layer CNN & 5.345  & 6.889 & 5.345  & 6.889 \\
	    No Peculiar Velocity & 4-Layer CNN & 5.066  & 6.569 & 5.066  & 6.569 \\
	    & 5-Layer CNN & 5.772  & 7.270 & 5.772  & 7.270 \\
	    & ResNet18 & 10.658 & 13.547 & 10.658 & 13.547  \\
	    & ResNet50 & 12.269 & 22.651 & 12.269 & 22.651 \\\hline
	    Normal Distribution & 2-Layer CNN & 8.826 & 11.152 & 7.510  & 9.553  & 0.975 & 668.717 \\
	    No Normalization & 3-Layer CNN & 9.015 & 11.811 & 7.747  & 10.258  & 0.991 & 718.068 \\
	    Peculiar Velocity & 4-Layer CNN & 9.029 & 11.315 & 7.729  & 9.969  & 0.976 & 678.941 \\
	    & 5-Layer CNN & 9.380 & 11.881 & 8.064  & 10.341 & 0.994 & 723.926 \\\hline
	    Normal Distribution & 2-Layer CNN & 7.522 & 10.781 & 6.056  & 9.198 & 0.946 & 643.889  \\
	    Normalized & 3-Layer CNN & 7.622 & 7.832 & 6.207  & 7.832 & 0.950 & 548.294  \\
	    Peculiar Velocity & 4-Layer CNN & 7.675 & 9.922 & 6.125  & 8.151 & 0.946 & 570.630 \\
	    & 5-Layer CNN & 7.580 & 10.766 & 6.154  & 9.257 & 0.923 & 647.995 \\
	    & ResNet18 & 9.855 & 28.635 & 8.647 & 28.192  & 0.968 & 503.732 \\
	    & ResNet50 & 15.247 & 55.557 & 14.280 & 55.331 & 1.001 & 1110.404 \\\hline
	    Uniform Distribution & 2-Layer CNN & 6.303 & 7.892 & 6.303 & 7.892 \\
	    No Normalization & 3-Layer CNN & 6.344 & 7.963 & 6.344 & 7.963 \\
	    No Peculiar Velocity & 4-Layer CNN & 7.771 & 9.732 & 7.771 & 9.732 \\
	    & 5-Layer CNN & 6.004 & 7.573 & 6.004 & 7.573 \\\hline
	    Uniform Distribution & 2-Layer CNN & 5.068 & 7.133 & 5.068 & 7.133 \\
	    Normalized & 3-Layer CNN & 4.535 & 5.770 & 4.535  & 5.770 \\
	    No Peculiar Velocity & 4-Layer CNN & 4.979 & 6.384 & 4.979 & 6.384 \\
	    & 5-Layer CNN & 4.767 & 6.119  & 4.767 & 6.119 \\
	    & ResNet18 & 8.913 & 12.101 & 8.913 & 12.101 \\
	    & ResNet50 & 12.653 & 20.902 & 12.653 & 20.902 \\\hline
	    Uniform Distribution & 2-Layer CNN & 8.608 & 10.868 & 6.870  & 8.776 & 1.041 & 474.140 \\
	    No Normalization & 3-Layer CNN & 8.564 & 10.940 & 6.846  & 8.854 & 1.044 & 443.632 \\
	    Peculiar Velocity & 4-Layer CNN & 8.991 & 11.517 & 7.338 & 9.495 & 1.045 & 474.418 \\
	    & 5-Layer CNN & 7.753 & 9.867 & 5.941 & 7.543 & 1.022 & 469.286 \\\hline
	    Uniform Distribution & 2-Layer CNN & 7.196 & 9.167 & 5.246  & 6.773 & 0.996 & 614.317 \\
	    Normalized & 3-Layer CNN & 7.070 & 8.873 & 4.901  & 6.227 & 1.001 & 619.802 \\
	    Peculiar Velocity & 4-Layer CNN & 7.313 & 9.193 & 5.157  & 6.777 & 1.004 & 664.653 \\
	    & 5-Layer CNN & 7.220 & 9.172 & 5.171 & 6.704 & 1.010 & 528.030 \\
	    & ResNet18 & 10.617 & 13.555 & 9.644 & 12.302 & 0.950 & 857.458 \\
	    & ResNet50 & 13.456 & 23.304 & 12.473 & 23.304 & 0.986 & 1045.560 \\\hline
	\end{tabular}
\end{table*}

\label{lastpage}
\end{document}